# UAVs Formation Flight under Multiple Obstacle and Communication Constraints

Hu Jin

College of Computer Science/Software School, University of South China, 421001,Hengyang Chnia

Hzz1988@usc.edu.cn

**Abstract**.

In order to improve the communication of the UAV network when the UAV formation in a complex military mission environment. This paper proposed formation reconfiguration strategy under complexities mission environment. We take the problem of multi-UAV formation reconfiguration transform into nonlinear inequality constraints optimization by mathematic approach. (i)design multi-objective multi-constraints optimization function for different task of UAVs and complexities mission environment, and get an optimal model.(ii) Using Pareto optimal theory transform the multi-objective optimization problem into single optimization problem.(iii)finally, we apply the primal-dual Newton interior point algorithm to process the convex optimal problem. In order to verify the high efficiency of this method, a series of simulation experiments are designed. The experiment shows that the method in this paper has the lowest energy consumption while maintaining the connectivity of the UAV formation network.

## 1 Introduction

With the rapid development and maturity of microcontroller design, optimization techniques, and control theory, it has greatly promoted the widespread use of UAVs in civil unclear military fields [1]. UAV a class of aircraft without human pilot onboard. Due to the multiple advantages of UAVs (such as flexibility, fast flight, crew endurance, etc.) in a variety of mission, it have been assigned high-risk mission so that can completely protect the human lives from danger [2].

In the highly information-based military battlefield, the UAVs execute a mission mode, which will transform the single automatic combat mode into multi-UAVs swarms to cooperative complete a task. The model of the UAV performance a mission will transform single UAVs into multiple UAVs cooperative complete. Multi-UAVs collaborative operations will inevitably require the friends of the UAVs through the sharing of information and the data links for unified decision-making, coordination of division of labor, to ensure multi-objective decision-making optimization. However, the implementation of the UAV collaborative task will faces many challenges, some key technical issue have not yet been solved. Such as large-scale UAV management and control, Multi- UAVs autonomous formation and flight, cluster awareness and situation sharing, cluster penetration and attack, and cluster combat mission control station. Generally speaking, UAVs swarms to perform a task must have autonomous formation capabilities. In recent years, formation control for unmanned aerial vehicle systems, have drawn significant attention from related research communities and become a research hotspot [3]. This paper is mainly focus on the problem of formation reconfiguration, which is one of the issue to be addressed in formation control.

In addition to research on UAVs formation flying, there have also been a number of studies on coordinating the behavior of multiple robots and spacecraft. While the application is different, the fundamental approaches to the coordination of multiple UAVs, robots and spacecraft are very similar [4].The formation control method is classified according to different criteria. From the measurement information perspective, formation control problem can be formulated displacement-based and distance-based methods. The method of displacement-based [5] is based on a common sense of orientation, agent measure the relative positions (displacements) of their neighbors with respect to a common reference frame. Then the agents directly control the relative positions to control their formation. This may increase the workload of the agent, and problem are complicated. At distance-based [5,6], agents measure the relative positions of their neighbors only with respect to their own local reference frames whose orientations are not necessary aligned with each other due to the absence of a common sense of orientation. But this approach is inefficiency to global asymptotic convergence. Kwang et.al [5] proposed a formation control strategy exploiting effectiveness of displacement-based approaches while using measurements that are assumed to be available in distance-based approaches. From the control mechanism perspective, formation control approach can be classified into consensus-based approaches [7-9], artificial potential function-based approaches [10-11], virtual structure approaches [12-15], leader-following approaches [16-20] and behavior-based approaches [21], methods of model-based formation control [22]. Yasuhiro et.al [7] adapt the method of consensus-based to study multi-UAV system formation control, but the method may also increase the workload of UAVs. Ranjith et.al [10] formulate an artificial potential function for planning path, and using sliding model control technique for designing a robust controller. Hamed et.al [15] proposed a formation control for unmanned aircrafts using virtual structure. They are establish a cross coupled sliding model to cope with uncertainties in the attitude measurement systems of the unmanned aircrafts and unmeasurable bounded external disturbance. They also consider the communication delays between two UAVs. Ali et.al [17] study the formation control for unmanned helicopters with hybrid three-dimensional by leader-following. Antonio et.al [18] proposed Leader-follower formation and tracking control of mobile robots along straight paths. Although the leader-following approach is easy to implement and understand, the formation is sensitive to the leader behavior, and disturbed or failed leaders affect their follower motion. The methods of virtual structure, leader-following and behavior are the special cases of consensus strategies. The behavior-based approach are not definition formation, and it is useful for multi objective missions such as target seeking, obstacle avoidance. Micheal et.al [22] studying the first and second order sliding-model controllers are proposed for asymptotically stabilizing the vehicles to a time-varying desired formation. There have anther interest approaches of formation control differential game approaches [23], this approach take the formation control problem formulated as a linear-quadratic Nash differential game by using the graph theory. From control structure perspective, the existing formation control approaches can be classified into centralized and distributed/decentralized methods [24-28]. This method may increase the communication burden on the entire network as the number of nodes increase.

## 2 RELATED WORK

Many researchers have investigated the formation control problems, but every few literatures focus on the reconfiguration issue of UAV formation. Reconfigurable control may be needed in case of failure in multi-UAVs ad hoc network, flight path constraints or even the total loss of the aircraft without impairing its mission. This work focus on formation reconfiguration problem. The concept of formation reconfiguration involves determining aircraft separation distance, position and orientation, identifying the process that optimally transform an initial formation configuration into a final configuration and identifying cooperative a desired final configuration. Several approaches have been applied to the formation reconfiguration problem. Duan et.al [29] proposed hybrid particle swarm optimization and genetic algorithm for multi-UAV formation reconfiguration. This paper discretization the input of each flight unit by a control parameterization and time discretization method at first. And then transform the multi-UAV formation reconfiguration problem into an optimal problem with dynamic and algebraic constraints. Finally, due to the multi-UAV formation reconfiguration in 3-D space is a complicated problem with strict constraints and mutual interference, they apply intelligent algorithms to solve the optimal value. Jiang et.al [30] proposed symplectic iterative numerical algorithm to obtain the optimal solution for the nonlinear receding horizon control strategy at each instant. Using a high-efficiency structure-preserving symplectic method in the iterations, and replace the optimal control problem by a series of sparse symmetrical linear equations. Ludwik at.al [31] study spacecraft formation reconfiguration by design an optimal control strategy of continuous/impulsive linear quadratic regulator (LQR) that combines continuous Lorentz force with impulsive thrusting. Park et.al [32] based on graph theory to establish a relation motion model involving communication topology of formation flying on a circular reference orbit. Zhou et.al [33] employing a novel predesigned desired velocity and an elaborate adaptive law, a finite-time coordination control scheme to drive all the spacecraft to implement the formation reconfiguration task in unkown obstacle environment. Hong et.al [34] modeling formation reconfiguration as a fuel optimal control problem, and using chebyshev pseudospectral method transform the above problem into a nonlinear programming problem. Ding et.al [35] introduce geometric relative orbit elements to depict formation configurations. They deal with non-convex constraints by combining external convexification iterations with Guass pseudospectral method. Because the complexity of mission environment, we formulate the multi-UAVs formation reconfiguration as multi-objective optimization under multi-constraints scenarios. Due to the traditional interior point algorithm has the difficult of search the initial feasible point and the algorithm convergence is poor [36]. In this paper, we using the primal-dual damp Newton interior point algorithm to solve the nonlinear inequality constraints optimization problem. The outline of this paper is as follows. In section 3, it is mainly describe the constraints and objective for multi-UAVs formation flight, and modeling an optimization model. In section 4, applying weight sum strategy transform the multi-objective optimization problems into single-objective optimization problems. Formulate the primal-dual Newton interior point algorithm for nonlinear inequality-constraints optimization problem. In section 5, it is mainly design a set of experiment to testify the method of this paper proposed. Concluding remarks are then provided in section 6.

## 3 Problem Description and Optimization Model

UAV formation reconstruction and formation maintenance are two problems to be solved by formation control. Formation maintenance can both refer to contraction or expansion of a formation while maintaining the same geometry, or to switching between different geometries. UAV formation re-construction refer to task redistribution according to the current flight status of the UAV and its neighbors under the circumstance of UAV encounter an obstacle or suffer from enemy attack. The realization of UAV formation reconstruction not only needs to consider the current flight state of the UAV in the original formation, but also consider the current environment and the formation of overall connectivity. UAVs state mainly refer to position, velocity, heading and flight path angle.

$$\dot{p}_{xi} = v_i \cos(\psi_i)$$
$$\dot{p}_{yi} = v_i \sin(\psi_i)$$
$$\dot{\psi}_i = \varpi$$

Where $p_i = [p_{ix}, p_{iy}]$ represent the ith UAV position at time t, $v_i$ is velocity, $\varphi_i$ is heading angle. The control input is $u_i(t) = [v_i, \varpi]$.

The ith UAV dynamic equation as follow

$$x_i(t+1) = f(x_i(t), u_i(t))$$

$$\begin{bmatrix} p_{xi}(t+1) \\ p_{yi}(t+1) \\ \varphi_i(t+1) \end{bmatrix} = \begin{bmatrix} p_{xi}(t) + v_{xi}(t)\Delta t \\ p_{yi}(t) + v_{yi}(t)\Delta t \\ \varphi_i(t) + \varpi_i \Delta t \end{bmatrix}$$

Because of the complexity and uncertainty of unmanned aerial vehicle (UAV) co-executing mission environment, it is necessary to consider multiple constraints to optimize formation control structure in order to make UAV formation efficiently and efficiently. In this paper, five constraint functions are designed to further optimize the above heterogeneous hierarchical formation control structure.

Terrain limited: The formation should fly above the rugged terrain and avoid collision with the mountain. Let the mountain as a regular pologons and set it high is h. the position of the ith UAV is $P_{iz}$. In order to avoid the UAVs at the edge of the formation collision with the mountain, this constrain can be depicted as

$$g_i^1(x_i(t), u_i(t)) = h - p_{iz}(t) \leq 0$$

Where the first term of above equation is express the formation on the current total energy consumed, the second term is the ith UAVs at time t.

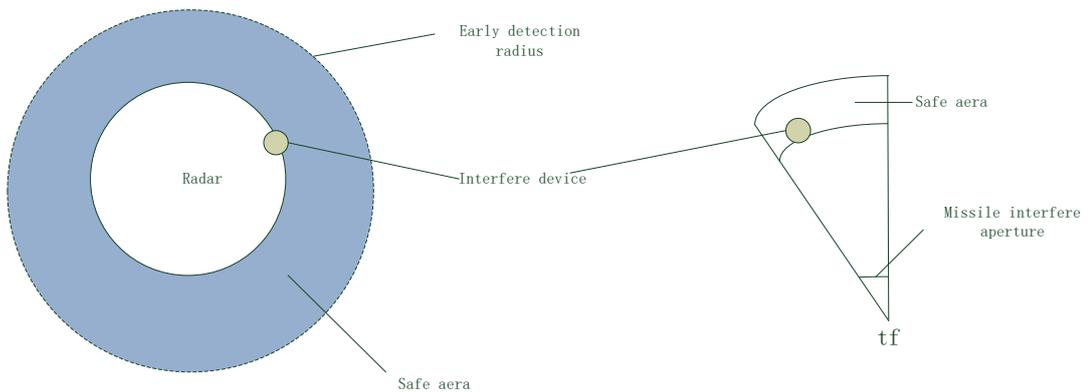

Assume the position and detection radius, in order to ensure the UAVs safety, it is need satisfy this condition

$$g_i^2(x_i(t), u_i(t)) = (R_{r_j}(t))^2 - \| p_i(t) - p_{r_j}(t) \|^2 \leq 0$$

Assume the parameters of $d_{m_j}, CS_{m_j}$ represent the safe distance and the safe cosine angle of missile, respectively. The missile position is $P_{tf}$, the interfere device position is $P_o$, the ith UAV position is $P_i$. In order to ensure that the UAV is not hit by missiles, the following conditions need to be met

$$d_{mj} = \| p_i - p_{tf} \| - \| p_o - p_{tf} \| \geq 0$$

$$CS_{mj} = \frac{(p_i - p_{tf})(p_o - p_{tf})}{\| p_i - p_{tf} \| \| p_o - p_{tf} \|}$$

$$g_i^3(x_i(t), u_i(t)) = CS_{m_j}(t) - \cos(\frac{\theta}{2}) \leq 0$$

Anti-collision constraints

Some UAVs are attacked when they are crossed the enemy's tight air defense zone, or the UAVs have to separate from the formation when they are encounter an obstacle. Then, the number of formation members is reduced and the network nodes is become sparse and the two UAV is disconnected. In order to enable mission data can be shared between UAVs, it is necessary to reconstruct the communication link to the UAV. The multi-UAVs formation can be modeled as an undirected weight finite graph $G=(V,E)$, where $V = \{v_1, v_2, ......, v_N\}$ is the set of all nodes (UAVs) and E is the set of all edges (links). An undirected graph implies that all the links in the network are bidirectional, hence, if node $v_i$ can reach node $v_j$ then the opposite is also true. A simple graph means that the cardinality of the set V and E is finite. Let n and m denote the number of nodes edges in the graph, respectively, i.e., $|V| = n$ and $|E| = e$, where $|\cdot|$ is the cardinality of the given set. Anti- collision between the UAVs is one of constraints to be considered for unmanned aerial vehicles swarms to cooperative tracking. This anti-collision force only work if the communication between the two considered UAVs is available, which is reasonable when the UAVs get closer each to other. Assumption that the minimum safety distance between UAVs is $d_{min}$, communication radius $R_i$ is calculated as follow:

$$R_i = \sum_{j \in adj\{i\}} \frac{1}{(\| p_i - p_j \| - d_{min})} \frac{p_i - p_j}{\| p_i - p_j \|}$$

In order to meet the above requirements, and there will be no collision between UAVs, the distance of UAVs should satisfy the following condition

$$g_i^4(x_i(t), u_i(t)) = d_{min} - \| p_i(t) - p_j(t) \|^2 \leq 0$$

Consider different UAV complete different task. We are design three different cost function for the reconnaissance UAVs, the radar interference UAVs and missile interference UAVs, respectively. Let the sets of N represent overall number of UAV in the formation, the number of reconnaissance UAVs is $n_1$, the number of missile UAVs is $n_2$ and the number of radar interference UAVs is $n_3$, $\{n_1 \cup n_2 \cup n_3\} \subset N$ Definition K is the task completion time.

Define the position of virtual leader aircraft at time t as follow

$$P_v(t) = (p_{vx}(t), p_{vy}(t), p_{vz}(t))$$

The position of the ith UAV whose role is to carry out the reconnaissance mission is defined as

$$p_i(t) = (p_{ix}(t), p_{iy}(t), p_{iz}(t))$$

The cost function corresponding to the ith UAV during the whole reconnaissance task is define as follow

$$F_i^1(x_i, u_i) = \sum_{t=1}^{K} (\|(p_v(t) - P_i(t))\|^2 + \|u_i(t)\|^2_{M_i})$$

Where $M_i$ is positive definition weighting matrix. The second term is used to normalized the original optimization problem and to guarantee that the optimal solution will not depart the true solution.

We are assume that the idea position and the current position of the missile interference aircraft at time t is $p_{l1}(t)$, $p_i(t)$, respectively. The cost function for missile UAV can be define as

$$F_i^2(x_i, u_i) = \sum_{t=1}^{K} (\|(p_{l1}(t) - P_i(t))\|^2 + \|u_i(t)\|^2_{M_i})$$

Same as above, the idea position and the current position of the radar interference UAV is $p_{l2}(t)$, $p_i(t)$, respectively. The cost function for radar missile UAV can be define as

$$F_i^3(x_i, u_i) = \sum_{t=1}^{K} (\|(p_{l2}(t) - P_i(t))\|^2 + \|u_i(t)\|^2_{M_i})$$

In summary, available UAV formation of the overall reconstruction optimization model as follows:

$$F(x, u) = \min [\underbrace{\overbrace{F_i^1(x_i, u_i)...}^{N} \underbrace{F_i^2(x_i, u_i)...}_{i \in n_2} \underbrace{F_i^3(x_i, u_i)...]}_{i \in n_3}}_{i \in n_1}$$

Subject to

$$H(u_i) = x_i(t+1) - f_i(x_i(t), u_i(t)) = 0$$
$$x_i(t) \in \Xi, u_i(t) \in \Theta$$
$$g_i^1(x_i(t), u_i(t)) \leq 0$$
$$g_i^2(x_i(t), u_i(t)) \leq 0$$
$$g_i^3(x_i(t), u_i(t)) \leq 0$$
$$g_i^4(x_i(t), u_i(t)) \leq 0$$
$$g_i^5(x_i(t), u_i(t)) \leq 0$$

Where $\Xi$ represents the state vector set, $\Theta$ is control vector set, respectively. Let

$$g_i(x,u) = \begin{bmatrix} g_i^1(x,u) \\ g_i^2(x,u) \\ g_i^3(x,u) \\ g_i^4(x,u) \\ g_i^5(x,u) \end{bmatrix} \leq 0$$

The optimal model can be rewritten as follows

$$F(x,u) = \min [\underbrace{F_i^1(x_i,u_i)...}_{i \in n_1} \overbrace{\underbrace{F_i^2(x_i,u_i)...}_{i \in n_2} \underbrace{F_i^3(x_i,u_i)...}_{i \in n_3}}^{N}]$$

s.t.
$$H(u_i) = x_i(t+1) - f_i(x_i(t), u_i(t)) = 0$$
$$x_i(t) \in \Xi, u_i(t) \in \Theta$$
$$g_i(x,u) \leq 0$$

## 4 Primal-dual Interior Point Algorithm
## 4.1 Pareto Optimal for Multi-objective Optimization

Prediction Considering the above optimization model is a multi-objective optimization. The idea of a solution for the above model can be clear, because a single point that minimizes all objective simultaneously usually does not exist. Consequently, the idea of Pareto optimality is used to describe solutions for multi-objective optimization. Pareto optimal means that at least one object get better without causing any one object to deteriorate. Typically, there are infinitely many Pareto optimal solution for a multi-objective problem. Thus, it is often necessary to incorporate the relationship between multiple objectives in order to determine a single suitable solution. Using the weighted sum method to solve the problem in the above formulation entails selecting scalar weights $\omega_i$ and minimizing the following composite objective function

$$F(x,u) = \sum_{i=1}^{n} (\omega_{1i} F_i^1(x_i,u_i) + \omega_{2i} F_i^2(x_i,u_i) + \omega_{3i} F_i^3(x_i,u_i))$$

s.t.
$$H(x,u) = 0$$
$$G(x,u) \leq 0$$

Where $G(x,u) = [g_1(x,u), g_2(x,u), ......, g_N(x,u)] \leq 0$.

Let the number of these three types of UAVs are equal to n, $n = \dfrac{N}{3}$

$$\sum_{k=1}^{3}\sum_{i=1}^{n} \omega_{ki} = 1$$

The weights represent the gradient of U in the above formulation with respect to the vector function F(x, u), show as follows

$$\nabla F = \begin{bmatrix} \dfrac{\partial F}{\partial F_1^1} & \cdots & \dfrac{\partial F}{\partial F_n^1} \\ \dfrac{\partial F}{\partial F_1^2} & \cdots & \dfrac{\partial F}{\partial F_n^2} \\ \dfrac{\partial F}{\partial F_1^3} & \cdots & \dfrac{\partial F}{\partial F_n^3} \end{bmatrix} = \begin{Bmatrix} \omega_{11} & \cdots & \omega_{1n} \\ \omega_{21} & \cdots & \omega_{2n} \\ \omega_{31} & \cdots & \omega_{3n} \end{Bmatrix}$$

The Pareto optimal solution for a given set of weight is found by determining where the U with the lowest possible value intersects the boundary of the feasible criterion space.

### 4.2 Primal-dual Newton interior point algorithm

The interior point method is a method of optimizing inside the feasible region. The interior point method has very good effect for solving the nonlinear inequality constrained optimization problem. The basic idea is to hope that in the process of optimization iteration, the feasible region always exists. Therefore, the initial point should be within the feasible region, and the "obstacle" should be set at the boundary of the feasible region so that the iteration points are the interior points of the feasible region. However, the traditional interior point method has some difficulties in finding the initial feasible point. In order to solve this problem, we propose an improved interior point method based on the interior path tracking method, which only requires that the relaxation variables and the Lagrange multipliers satisfy the condition of greater than or less than zero in the optimization process. Instead of having to solve the problem in the feasible domain, the calculation process is simplified.

In this section, we formulation the primal-dual Newton interior point method for the nonlinear inequality constraints optimization problems. The basic principle of the primal-dual interior point method is that the relaxation variable is introduced to transform the function inequalities constraints in the mathematical model into an equation constraints, and the relaxation variable can only satisfy the simple great than zero condition at first. And then, in order to make the solution always within the feasible area, we introduce a perturbed function into the original objective function, and eventually get a barrier function. Finally, the original optimization problem is transformed into a nonlinear equality constraints optimization problem by the above steps. So the optimization problem can solved by the Lagrange multiplier method, directly. The Lagrange function is satisfy the perturbed KKT sufficiency/ necessary conditions. Finally, we will be to consider the damped Newton method applied to the perturbed KKT conditions.

The inequalities constraints nonlinear optimization problem can be transformed to an equalities constraints problem, with a nonnegative slack variable

$$F(u) = \sum_{i=1}^{n} (\omega_{1i} F_i^1(u_i) + \omega_{2i} F_i^2(u_i) + \omega_{3i} F_i^3(u_i))$$

s.t.

$$H(u) = 0$$
$$G(u) + s = 0$$

Where s is the slack variable.

Considering the Lagrange multiplier can be used to solve the equality constraints problem, we can define a Lagrange function for the above objective and equality constraints

$$L(u, \lambda, w, s) = F(u) + \lambda^T (G(u)+s) + w^T H(u) - \mu \sum_{j=1}^{m} \log(s_j)$$

The perturbed Karush-Kuhn-Tucker condition for this problem is

$$F(u, \lambda, w, s) = \begin{bmatrix} \nabla_u U(u) - \nabla_u G(u)^T \lambda + \nabla_u H(u)^T w \\ G(u) + s \\ H(u) \\ \lambda S e - \mu e \end{bmatrix} = 0, \quad (\lambda, s) \geq 0$$

Following the theory in [35], the standard Newton's methods assumption to our optimization problem are these:

(1) Existence. There exists $(u^*, \lambda^*, w^*)$, solution to problem (18) and associated multiplier, satisfying the KKT conditions.

(2) Smoothness. The Hessian matrices $\nabla^2 U, \nabla^2 H_i(u), \nabla^2 g_i(u)$ for all i exist and are locally Lipschitz continuous at $u^*$.

(3) Regularity. The set $\{\nabla H_1(u^*), \ldots, \nabla H_N(u^*)\} \cup \{\nabla g_i(u^*)\}$ is linearly independent.

(4) Second-order sufficiency. For all $\eta \neq 0$ satisfying $\nabla H_i(u^*)^T \eta = 0, i = 1, \ldots, N$ and $\nabla g_i(u^*)^T \eta = 0$. We have $\eta^T \nabla_u^2 L(u^*) \eta > 0$.

(5) Strict complementarity. For all i, $\lambda^*_i + g_i(u^*) > 0$

The Jacobian matrix $F'(u, \lambda, w, s)$ of $F(u, \lambda, w, s)$ as follow

$$F'(u, \lambda, w, s) = \begin{bmatrix} \nabla_u^2 L & -\nabla_u^T G & \nabla_u^T H & 0 \\ \nabla_u^T G & 0 & 0 & I \\ \nabla_u^T H(u) & 0 & 0 & 0 \\ 0 & I & 0 & -I \end{bmatrix}$$

Where $\nabla^2 L = \nabla^2 L(u, \lambda, w, s)$.

The sub-matrix of Jacobian matrix $F'(u,\lambda,w,s)$ are corresponding to equality-optimal constraints sub-problem, form the theory of the equality-constrained optimization, we can see that condition (3) and (4) of the standard Newton's are equivalent to the non-singularity of the matrix

$$F(u,\lambda,w) = \begin{bmatrix} \nabla_u^2 L & -\nabla_u^T \hat{G}(u) & \nabla_u^T H(u) \\ \nabla_u^T G(u) & 0 & 0 \\ \nabla_u^T H(u) & 0 & 0 \end{bmatrix}$$

It is not different to see that the $F'(u,\lambda,w,s)$ is non-singularity and equivalent to strict complementarity.

The primal-dual Newton interior-point method the nonlinear optimization problem. At the kth iteration, let

$$v_k = (u_k, \lambda_k, w_k, s_k)$$

Obtain the perturbed Newton correction

$$\Delta v_k = (\Delta u_k, \Delta \lambda_k, \Delta w_k, \Delta s_k)$$

Corresponding to the parameter $\mu_k$, as the solution of the perturbed Newton linear system

$$F'_\mu(v_k)\Delta v_k = -F_\mu(v_k)$$

$$\begin{bmatrix} \nabla_{u_k}^2 L & -\nabla_{u_k}^T G & \nabla_{u_k}^T H & 0 \\ \nabla_{u_k}^T G & 0 & 0 & I \\ \nabla_{u_k}^T H(u) & 0 & 0 & 0 \\ 0 & I & 0 & -I \end{bmatrix} \begin{bmatrix} \Delta u_k \\ \Delta \lambda_k \\ \Delta w_k \\ \Delta s_k \end{bmatrix}$$
$$= -\begin{bmatrix} \nabla U(u_k) - \nabla G(u_k)^T \lambda_k + \nabla H(u_k)^T w_k \\ G(u_k) + s_k \\ H(u_k) \\ \lambda_k S_k e - \mu_k e \end{bmatrix}$$

If our choice of steplengths are $\alpha_u, \alpha_\lambda, \alpha_w, \alpha_s$, we constructed the expanded vector of steplengths

$$\alpha_k = (\alpha_u,......,\alpha_u,\alpha_\lambda,......,\alpha_\lambda,\alpha_w,......,\alpha_w,\alpha_s,......,\alpha_s)$$

Where the frequencies of occurrences of the steplengths are $m, n, p, p, p$ respectively. Now we let

$$\Lambda_k = diag(\alpha_k)$$

$\Lambda_k$ is a diagonal matrix with diagonal $\alpha^k$. Hence, the subsequent iterate $v_{k+1}$ can be written as

$$v_{k+1} = v_k + \Lambda_k \Delta v$$

For global convergence consideration, a merit function $\phi(v)$, that measures the progress towards the solution $v^* = (u^*, \lambda^*, w^*, s^*)$ should be used. The merit function used for the linear search is the square $l_2$-norm of the residual.

$$\phi(v) = \| F(v) \|_2^2$$

Algorithm1 primal-dual interior-point algorithm

Step0. Let $v_0 = (u_0, \lambda_0, w_0, s_0)$ be an initial point satisfying $(\lambda_0, s_0) > 0$. For $k = 0, 1, 2, \ldots$, do the following steps

Step1. Test for convergence

Step2. Choose $\mu_k > 0$.

Step3. Solve the linear system $F_\mu^{'}(v_k) \Delta v_k = -F_\mu(v_k)$.

Step4. Compute the quantities

$$\hat{\alpha}_s = -1 / \min((S_k)^{-1} \Delta s_k, -1)$$
$$\hat{\alpha}_\lambda = -1 / \min((\lambda_k)^{-1} \Delta \lambda_k, -1)$$

Step5. Choose $\tau_k \in (0,1]$ and $\alpha_p \in (0,1]$ satisfying

$$\phi(v_k + \Lambda_k \Delta v) \leq \phi(v_k) + \beta \alpha_p \nabla \phi(v_k)^T \Delta v_k$$

For some fixed $\beta \in (0,1)$, where $\Lambda_k = diag(\alpha_k)$ with the steplength choices

$$\alpha_u = \alpha_p, \alpha_w = \alpha_p$$
$$\alpha_s = \min(1, \tau_k \hat{\alpha}_s), \alpha_\lambda = \min(1, \tau_k \hat{\alpha}_\lambda)$$

Step6. Set $v_{k+1} = v_k + \Lambda_k \Delta v_k$ and $k \leftarrow k+1$. Go to step1.

## 5 Simulation

The A UAV formation of six UAV is considered as an example in order to approach actual unperturbed relative UAV formation flying motion on a smooth reference path. This formation include three radar interference UAV, two reconnaissance UAV and one missile interference UAV. The initial configuration is coupled double regular tetrahedrons, as depicted in fig.1, where the length of each edge is 1km and the ideal flight path of UAV1 is specified to coincide with the reference path. An example communication topology for specified formation is illustrated in fig.2. The initial position and technical parameters as table.1. There exist three enemy radar, one enemy guided missile and some mountains (or high buildings) in the mission environment.

| UAV | Type | Position(km) | Velocity (min, max, $\Delta v$) | Yaw |
|---|---|---|---|---|
| V1 | reconnaissance | (0,0,500) | [80,15, ±5 ] | ±2⁰ |
| V2 | reconnaissance | (-500,-1500,-500) | [80,15, ±5 ] | ±2⁰ |
| V3 | Radar interfere | (-500,0,-1000) | [80,15, ±5 ] | ±2⁰ |
| V4 | Radar interfere | (-500,500,1500) | [80,15, ±5 ] | ±2⁰ |
| V5 | missile interfere | (-1000,-1000,0) | [80,15, ±5 ] | ±2⁰ |
| V6 | Missile interfere | (-1000,-1000,1000) | [80,15, ±5 ] | ±2⁰ |

We set the maximum number of iterations C=100, the iteration terminate threshold value is $\xi=0.01$, minimum security distance $d_{\min} = 40$. The interfere aperture of radar interference UAV is $\theta = 30^0$. Consider the location of the deployed radar, we let the V3, V4 to interfere the radar 1, radar 2, V5, V6 to interfere the missile 1, missile2, respectively. Setting the time of simulation is 200s.

Table2 Simulation parameters

| Enemy device | Position(km) | Relative parameters |
|---|---|---|
| Radar 1 | [500,3000,0] | Initial detection distance > 5000m |
| Radar 2 | [1000,3500,0] | Initial detection distance > 5000m |
| Guide missile | [4000,0,0] | Interfere angle 10 |
| Guide missile | [5000,0,0] | Interfere angle 10 |
| Mountain | | Maximum height=1000m |
| UAV | | Maximum transmission distance=10km |

In order to estimate the efficiency of multi-UAV formation reconfiguration, we define the probability of radar detected $P_r$ and probability of missile detected $P_m$, respectively.

$$p_r = \frac{the\ time\ of\ radar\ exposure}{total\ simulation\ time}$$

$$p_m = \frac{the\ time\ of\ radar\ exposure}{total\ simulation\ time}$$

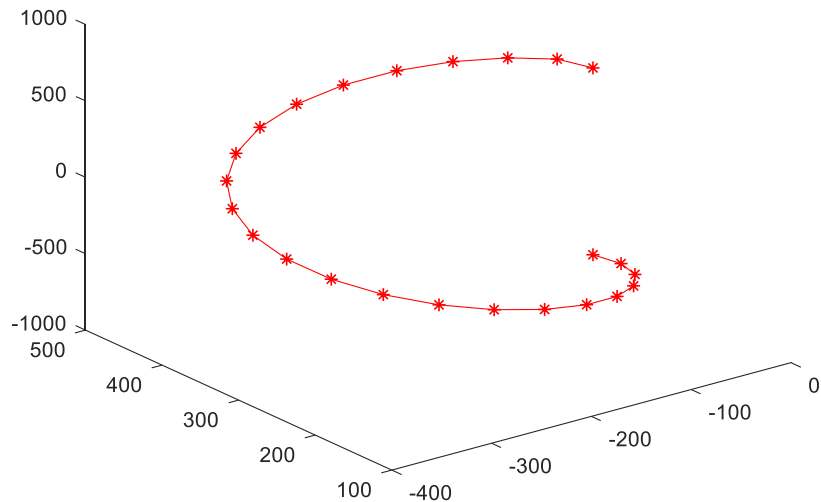

Fig. 2 UAV trajectory

Fig.2 shows the comparison of energy efficiency, total energy consumption and total throughput obtained by different schemes during the process of increasing the number of real-time users from 1 to 5, in which represents the change in energy efficiency as the number of real-time users increases, shows the change in total energy consumption with the increase of the number of real-time users, shows the change in total throughput with the increase of the number of real-time users. From fig. 2, we can see that the energy efficiency, the total energy consumption of UAV and the total throughput decrease with the increase of the number of real-time users.There are three reasons for this consequence; first,as the number of real-time users increase, the bandwidth resources allocated to UAV MEC decrease, thereby reducing the amount of computing task offloading; secondly, the total bandwidth resources of the communication system are fixed and increasing the number of real-time users leads to a decrease in the bandwidth allocated to non-real-time users in downlink; Finally, in the case of limited bandwidth resources, in order to enable UAV to cover real-time users at all times and meet the data rate requirements of all users in downlink, the trajectory of UAV is flatter, so that the energy consumption of UAV decrease with the number of real-time users increases. experiment.

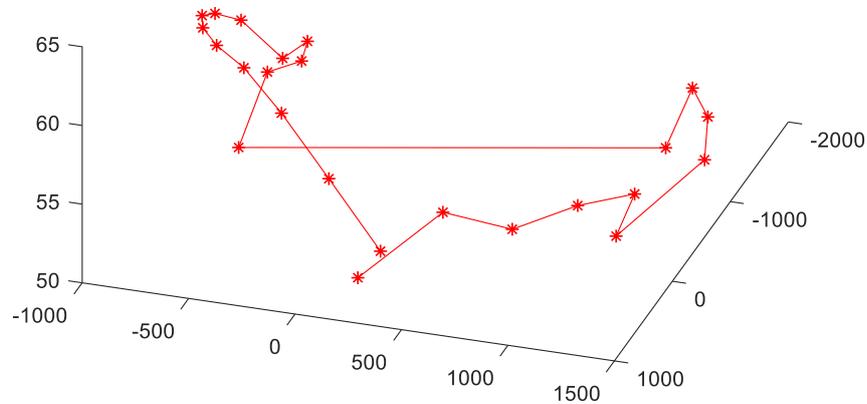

Fig .3 UAV trajectory

Next, we compares the UAV 3D trajectory of different schemes. The curve in fig.3 is the UAV trajectory obtained by the proposed scheme and the scheme with goal of throughput maximization, respectively. As show in fig.3, the increment of the UAV of the proposed scheme is smaller than that in the scheme with goal of throughput maximization. There are two reasons for this, on the one hand, in the scheme aiming at throughput maximization, the UAV usually adjusts its trajectory to maximize the instantaneous data rate, so as to shorten the distance between it and ground user as much as possible.

## 5 Conclusion

In this paper, we investigate the UAV trajectory joint with resource allocation optimization in MEC coordinated multi-UAV assisted communication networks. In the scenarios where different

UAVs perform different tasks, multiple UAVs compete for network bandwidth resources, and UAVs interfere with each other's communication, we deeply discuss the influence of UAV trajectory, user communication scheduling, bandwidth allocation, and UAV transmission power on the energy efficiency of the communication network. Since the real-time trajectory position of UAV directly affects the resource allocation and the QoE of user, a joint optimization model of UAV 3D trajectory and resource allocation aiming at maximizing the energy efficiency is designed. Because the optimization model is composed of fractional objective function and non-convex constraints, the problem is difficult to solve. Therefore, this paper is divided into three steps to solve the problem.